\documentclass[aps,reprint,twocolumn]{revtex4-1}
\usepackage{amsmath}
\usepackage{graphicx,color}
\usepackage{booktabs}
\usepackage{datetime}
\definecolor{navyblue}{rgb}{0.0,0.0,1}
\usepackage{xcolor}
\usepackage{epstopdf}
\usepackage{wrapfig}
\usepackage{ragged2e}
\usepackage{epsfig}
\usepackage{slashed}
\usepackage{caption}
\usepackage{subcaption}

\begin{document}
\title{Wake effects of a  stationary charged grain in streaming magnetized ions}


\author{Sita Sundar}

\affiliation{Department of Aerospace Engineering, Indian Institute of Technology Madras, Chennai - 600036, India}
\begin{abstract}
A systematic numerical study of wake potential and ion density distribution of a single grain in streaming ions 
under the influence of the magnetic field applied along flow is presented. 
Strong magnetic field introduces ion focus depletion behind grain facilitating the entrance of  electrons far away in the downstream towards the grain. 
It is shown that the magnetic field suppresses the amplitude of wake potential and modifies the ion density distribution substantially.
The wake peak potential and  position characteristics, and 
density distribution of plasma constituents  in the 
presence  of magnetic field and charge-exchange collisions  for the subsonic, sonic, and supersonic regime is also delineated.   In the subsonic regime, simulations demonstrate the accumulation of ions near  dust grain in the transverse direction while complete suppression of oscillations in the transverse direction  takes place for sonic and subsonic regime.
\end{abstract}
\maketitle
\section{Introduction}
Dusty plasma is ubiquitous in nature and laboratory plasmas i.e. spokes of Saturn rings, interplanetary dust, charged ice particles near moon, 
noctilucent clouds in Earth's atmosphere, Fusion devices etc. are various manifestations of dusty plasmas. Due to their heavy mass and the ability
to acquire high charge,  presence of dust particles in plasma makes them responsible for a variety of new novel phenomena which has been widely reported~\cite{Morfill:RMP2009,Melzer:WVV2008, Bonitz:Book2010}.
 Recent advances in scientific tools and technology has led to a surge in the interest of scientific community towards phenomena in dusty plasma regime which were hidden hitherto.
One among these is the study of impact of magnetic field on the dynamics of grain in streaming magnetised plasma. Magnetic field influences the behavior of charged plasma particles i.e. ions, electrons, and hence affects the overall dynamics of the system. It is well known that the influence of magnetic field introduces anisotropy in plasmas.  Dust particulates are heavy and it takes comparatively higher strength of magnetic field to make the grain magnetized.   It is pertinent to ask about the role of magnetic field on dusty plasma phenomena especially the exciting wake field features reported for the case of grains in streaming ions in the sheath region. Recently some experiments are being conducted and a few theoretical/numerical work in this regard has been reported~\cite{Edward:IEEE2013,Edward:POP2016,Miloch:JPP2014,Joost:PPCF2015}. 

The study of grain in magnetized ion flow began with the work by Nambu et.al.~\cite{Nambu:PRE2001}. They  provided the impact of magnetic field on the wake features for grain in streaming ions. In their paper, using analytical methods, they described that the role of magnetic field is to damp the strength of wakefield due to the reduction in ion overshielding around grain. In another work, Shukla et. al.~\cite{Shukla:PLA2001} presented the effect of ion polarization drift on  dynamical potentials and shielding  for magnetized plasmas.
Samsonov et.al.~\cite{Samsonov:NJP2003} discussed the impact of magnetic field on complex plasma from a different viewpoint. They demonstrated the levitation and agglomeration of magnetic grains in a complex plasma, and also envisaged the possibility of magnetically induced plasma crystal formation. Interest of wider scientific community in magnetized dusty plasma stemmed from these works.

Around the same time, Yaroshenko et.al.~\cite{Yaroshenko:NJP2003} described the fine details of mutual interactions of magnetized particles in complex plasmas. They presented that the dipole short-range force is the reason behind  the formation of field-aligned individual particle containing chains often observed in experiments.
Carstensen et.al.~\cite{Carstensen:PRL2012} presented the description of the inter-particle forces mediated by ion wakes in the presence of a strong magnetic field aligned along the ion flow.  Their observation was a decay in the interaction force with increasing magnetic field strength. They provided the reasoning for the decay at a critical parameter range where the ion cyclotron frequency is higher than the ion plasma frequency.

Recent preliminary results from MDPX experiments by Thomas et.al.~\cite{Edward:IEEE2013, Edward:POP2016} has revived the interest and aim of dusty plasma physicists towards the study of grain in magnetic field. They discussed the formation of ordered structures, properties of dust density waves, and filament generation with and without magnetic field.  The work relevant to the physics of wake formation in weakly and strongly magnetized plasmas is yet to be explored. 
One numerical work regarding wake formation for grain in streaming magnetized ions has been presented very recently by Miloch et.al.~\cite{Miloch:JPP2014} where they demonstrated that the wake size and strength can be significantly affected by the presence of the magnetic field for both stationary and streaming ions.

Studies by Joost et.al.~\cite{Joost:PPCF2015} were done for grain in magnetized ions using Linear Response (LR) formalism and they reported damping of wake-field with increasing magnetization.  
Note that  they studied the wake-field for a moving grain in stationary ions. This is not equivalent to the case of stationary grain in streaming ions because here we have magnetic field which introduces anisotropy. The velocity components are not isotropic in all the three directions and one should not substitute the dielectric function for stationary grain in streaming magnetized ions with streaming grain in stationary ions.
Nevertheless, the study served as a fruitful attempt to explore the effect of a magnetic  field on the screening potential.
In a recent work,  study for grain wakefield and induced charge density and other parametric dependence has been done~\cite{Ludwig:EPJD2017}  wherein they compared LR results with Particle-in-Cell (PIC) Simulations and have shown that the results have qualitatively similar characteristics. 

\begin{figure}
 \includegraphics[scale=0.45, trim = 9.0cm 6cm 1cm 5cm, clip =false, angle=0]{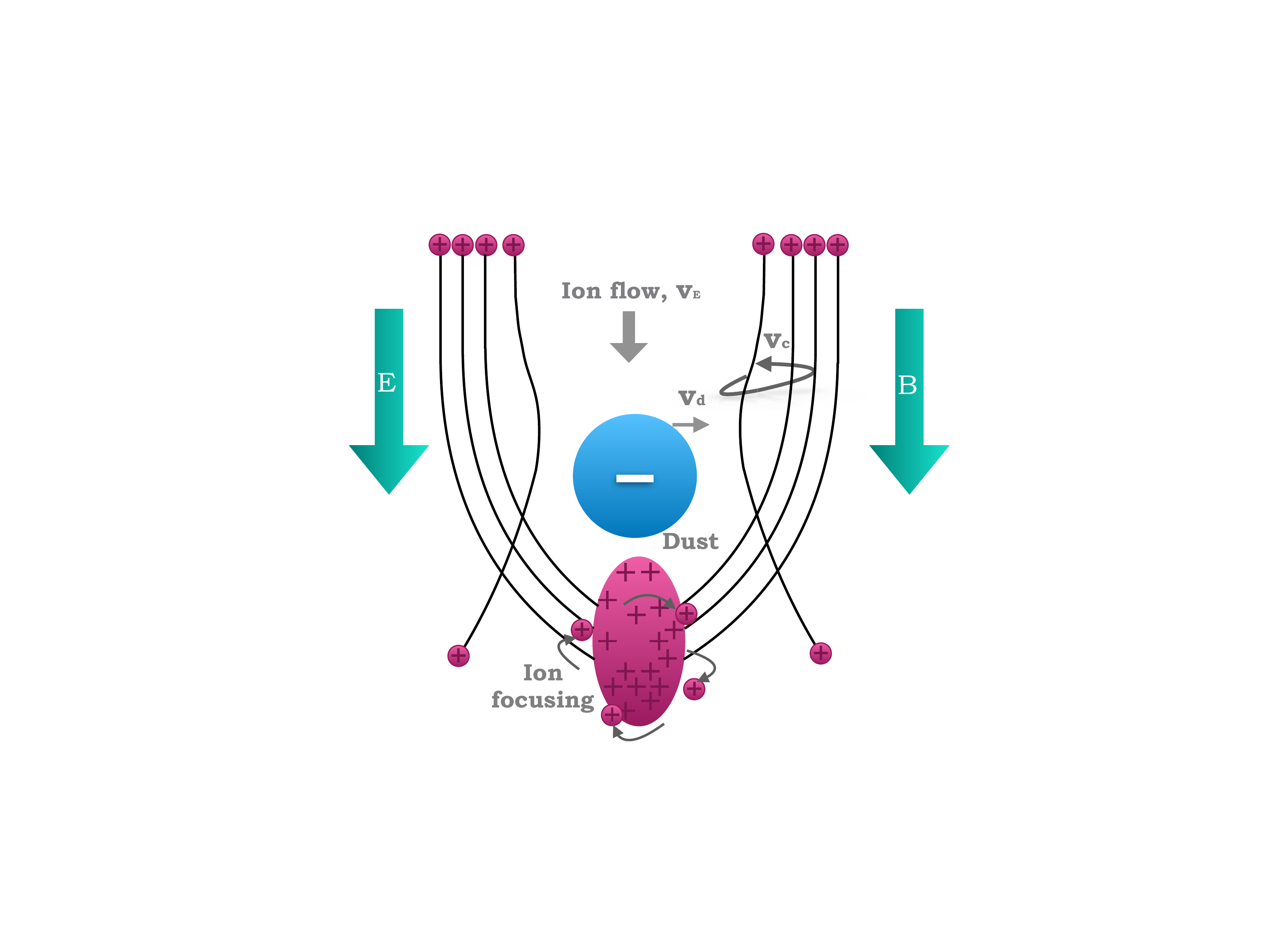} 
\caption{Schematic  depicting the ion streaming past grain and focusing downstream eventually leading to wake formation  for non-zero magnetic field applied parallel to the ion flow.}
\label{fig:Figure1}
\end{figure}

In the present work, we are going to study the impact of magnetic field on the wake-field formed downstream a single grain due to streaming ions, and their eventual impact on
density distribution around grain. A schematic of the system considered is presented in Fig.~\ref{fig:Figure1}. Here, we depict the streaming of magnetized ions past grain and focusing downstream grain. The force due to magnetic field alters the configuration substantially by introducing gyration and crossed magnetic drifts.
%
We know that ion focusing takes place downstream grain for the case of a stationary grain in streaming ions. 
But if one applies magnetic field along the flow direction, the streaming ions get magnetized (and dust also depending on the strength of the magnetic field applied) and their behavior along parallel and transverse directions are altogether different. This symmetry breaking can play a crucial role in the ion focusing downstream grain and  lead to significant modifications in  the grain-plasma dynamics as well as wake effects.

 In the presence of magnetic field, the ions passing far away from the grain (i.e. with large impact parameter) stream the way they used to in the absence of magnetic field nevertheless with gyration. Some of these streaming ions move past grain  eventually leading to ion focusing and wake effects in the downstream.  For ions streaming nearby grain, due to the influence of magnetic field and field due to negatively charged grain and wake potential, one comes across three different motions.  One among them is flow to due to  $dc$  field in the sheath which leads to accelerating flow, $v_E$. The second one is $v_c$, the gyrating motion around magnetic field. The third important force flow is $v_d$ resulting due to cross drift of  $E_{grain}$ or wake field and $B$ field (since external $E$ and $B$ applied are along the flow, they don't lead to any $E\times B$ drift). This $v_d$ shifts the ion focus behind grain and creates an ion depletion region through which electrons far away in the downstream region stream towards the grain. It has been observed in the present work that this ion depletion region has a strong dependence on the amplitude of magnetic field as well as streaming ion speed.
 
Here, the fact which also needs to be  emphasized is the role played by collisions. In the absence of collisions, particles do not diffuse in the perpendicular direction, rather they keep gyrating about the same line of force. The possibility of drift on these particles across $\vec{B}$ could be incurred due to electric fields or gradients in $\vec{B}$.  However, collisions assist the  particle diffusion  across $\vec{B}$ by a  random-walk process.  
Whenever an ion suffers a collision with a neutral atom, it undergoes a change in direction.
 It keeps gyrating about the magnetic field in the same direction, but with  change in phase of gyration and the guiding center shifts its position.
 The particles can then diffuse in the direction opposite to the density gradient. 
The magnitude of Larmor radius $r_L$ is the governing factor in determining the scale length of the random walk. This supplies us with the knowledge that one needs to maneuver the strength of $\vec{B}$ to control the diffusion across $\vec{B}$, and hence, other manifesting magnetized dusty plasma phenomena like coherent structure and wake formation.

The outline of the paper is as follows. In  Sec.II, we introduce the simulation scheme utilized and present  the description of methodology. 
 In Sec.III, we present the systematic results regarding the impact of magnetic field on the grain in streaming ions.
Finally, we 
present a summary and conclusion in section IV followed by acknowledgments in section V.

\section{Numerical Details}

The equation to delineate the ion dynamics in six-dimensional phase space in the presence of the self-consistent electric field $-\nabla \phi$,  an optional external force {\bf D}~\cite{Hutch:POP2013} and an external magnetic field is given by
\begin{equation}
 {m_i} \frac{d \mathbf{v}}{dt} = e \left[- \nabla {\phi} + \mathbf{v}  \times \mathbf{B} \right]+ \mathbf{D} .
\end{equation}
Here, $\bf{B}$ denotes the applied magnetic field and is aligned along the direction of ion flow in our simulations.
For  the {\it shifted Maxwellian distribution}, this extra force $\bf D$ is zero for most of our simulations, and ions are driven solely by a flow of neutrals. 
Simulation is performed with three-dimensional Cartesian mesh, oblique boundary, particles and thermals in cell (COPTIC) code~\cite{Hutch:POP2011}. COPTIC is a hybrid PIC code in a sense that electron dynamics are governed by the Boltzmann description, $n_e = n_{e \infty}\exp(e\phi/T_e)$, whereas ion dynamics are considered in six-dimensional phase space in the presence of the self-consistent electric field and optional external fields.  

The simulation set-up is similar to the one considered in our recent paper~\cite{Sundar:POP2017} except that we have an extra magnetic field aligned along the flow to take care of.
Further numerical detail and fundamentals can be gleaned from the paper with descriptions about COPTIC~\cite{Hutch:POP2011, Hutch:POP2013}.


To perform simulations, we chose a cell grid  of $64\times64\times128$ with more than 60 million ions and grid side length of $8\times8\times20$ Debye lengths. We performed few simulations with grids of even higher resolution and non-uniform mesh spacing to resolve the dynamics in the near neighborhood of the grain~\cite{Hutch:POP2011}.
Normalizations for the  length scale, velocity  and other physical variables  follow the standard normalization described in the  paper by Hutchinson et al.~\cite{Hutch:POP2011}, i.e., the space coordinate is normalized as $r\rightarrow r/r_0$,  velocity as $v \rightarrow v/c_s $, and potential as $\phi \rightarrow \phi/(   T_e/e)$, where $r_0=( \lambda_{De}/5)$ is the normalizing scale length and $c_s$ is unity in normalized units. The collision frequency $\nu$ is normalized in the time units as $\nu/(c_s/r_0) \sim 0.2(\nu/\omega_{pi})$, where $\omega_{pi}$ is the ion plasma frequency and the Debye length is fixed at $5 r_0$.  We  define ion Mach number $M$  in terms of the thermal Mach number $M_\text{th}$ as $M=v_d/c_s=\sqrt{T_i/T_e}\, M_\text{th}$, where $c_s=\sqrt{T_e/m_i}$ is the ion sound speed and $M_\text{th}=v_d/v_\text{th}$ is the thermal Mach number. Time-advancement of the simulation run till it reaches steady-state which is usually 1000 time-steps for the cases considered herein.  The analytical radius for point-charge sphere is chosen as $r_a=0.1 \lambda_{De}$. All the simulation parameters are summarized in Table~\ref{table:TABLE I}.
%

Incorporation of collisions are carried out in accordance to Poisson statistical distribution with fixed velocity-independent collision frequency i.e. similar to the BGK-type collisions. We have considered mainly the charged-exchange collision which is the dominant one in the system considered here, and is incorporated in the code by mutual interchange of velocity of the colliding ion with that of the velocity of neutral chosen randomly from the neutral velocity distribution. Magnetic field considerably modifies the dynamics by changing the wakefield potential amplitude, number of wake peaks behind grain and pattern of oscillations behind the grain, and  is discussed in sec.~\ref{sec:Res}.


\begin{table}[h]
\caption{Detailed list of the simulation parameters.}. 
\label{param_table}
\setlength{\tabcolsep}{7 pt}
\hspace*{-1cm}
\begin{tabular}{l l  }
\toprule[0.80pt]
\hline
\hline
 Magnetization, $\beta$ & 0.0-2.0\\
 Temperature ratio, $T_e/T_i$ & 100 \\
  Mach Number, $M$ & 0.5 - 1.5 \\
  Collision frequency, $\nu/\omega_{pi}$  & 0.002  \\
  Electron Debye length, $\lambda_{De}$  & 5 \\
  grid size, & $64 \times 64 \times 128$ \\
  number of particles, & $60 \times 10^6$\\
  total number of time steps, & 1000 \\
  Grain potential, $\phi_a$  & 0.05-0.2 \\
  Time-step, $dt$ & 0.1 \\
  Normalized grain charge, $\bar{Q}_d$ & 0.01\\ 
 
\hline
\hline
\bottomrule[0.80pt]
\end{tabular}
\label{table:TABLE I}
\end{table}
\section{Results}\label{sec:Res}
We have performed a systematic study of the wake potential as a function of both, the Mach number and the magnetization in the presence of 
 collisions. The electron-ion temperature ratio was fixed at $T_e/T_i =100$.  To delineate the differences in the wake potential for the unmagnetized and magnetized ion cases,  we performed simulations for parameters in the range $M=v_d/c_s=0.5-1.5$ and $\beta$ = 0.0 - {2.0} for $\nu=0.002$, where $\beta$ represents the magnetization parameter defined as the ratio of ion cyclotron frequency and ion plasma frequency, $\beta=\omega_{ci}/\omega_{pi}$.
\begin{figure*}
\includegraphics[height=18cm,width=15cm, trim = 6.9cm 6.5cm 1.9cm 5.9cm, clip =true]{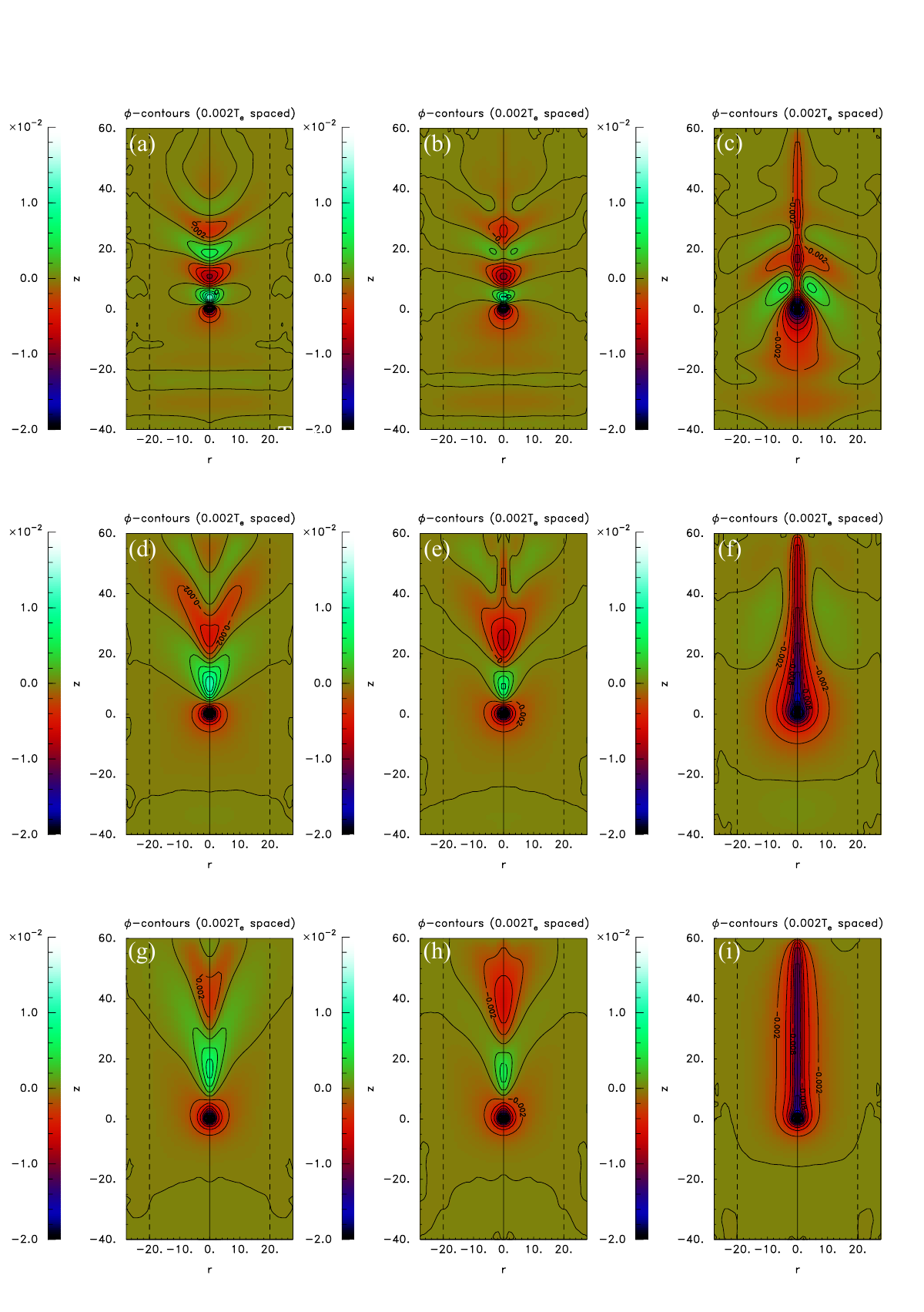}
    \caption{Wake potential contours $e \phi / T_e$, averaged over the azimuthal angle, for various strengths of magnetic field :
    (a) $\beta=0.05$; (b) $\beta=0.1$, and (c) $\beta=1.0$ (from left to right) with streaming velocity $M=0.5$ (top row), $M=1.0$ (middle row), and $M=1.5$ (bottom row) for the shifted Maxwellian case.
    }\label{fig:Figure2}
\end{figure*}
\subsection{Interplay of magnetization and ion streaming speed on wakefield potential}
Impact of ion streaming speed and ion-neutral charge-exchange collision has been widely reported in earlier works~\cite{Hutch:POP2013, Sundar:POP2017, Ludwig:NJP2012}.  
At first, we investigate in detail the impact of magnetzation on wakefield produced downstream grain due to ion focusing or depletion.  In Fig.~\ref{fig:Figure2}, we present wake potential contours, for various magnetizing stregths, for the shifted Maxwellian distribution with moderate to high streaming speeds. Similar to the observations made in linear response calculations~\cite{Joost:PPCF2015}, here also, we  observe that the role of magnetization is to reduce the amplitude of wakefield oscillations behind the grain.
We start with the regime of very small magnetization, $\beta=0.05$, Fig.~\ref{fig:Figure2} (left column). The magnetization strength is meagre and the difference from unmagnetizaed case, cf. Fig.~\ref{fig:Figure5}~\cite{Sundar:POP2017}, is inconspicuous and its wake potential profile is similar to that of the unmagnetized case for all the Mach numbers considered. However, due to the presence of magnetic field one can notice  a very slight modification  appearing in the wake potential at the far end of the grain  in the downstream.

 As we further increase the magnetic field strength, we see the growing impact of magnetization on the wake potential profile. For moderate magnetization, $\beta=0.1$, Fig.~\ref{fig:Figure2} (middle column), streaming of electrons in the upstream starts at the far end of the grain in the downstream and traverses through the ion focus. The impact of magnetization is stronger for higher ion streaming speeds. It can be observed that the role of streaming speed is to elongate the wake oscillation wavelength,  and that of the magnetic field is to suppress the wake amplitude.
 For smaller value of applied magnetic field, the depletion of ion focus starts from the far end of the wake oscillations downstream grain and this depletion region moves slowly towards the grain with increasing magnetic field strength eventually facilitating the entrance for faraway electrons in the downstream to move upstream towards the grain.

 At even higher magnetization strengths $\beta=1.0$, Fig.~\ref{fig:Figure2} (right column), the change in wake feature is prominent even in subsonic regime. Here, electrons are able to propagate upstream with less obstruction through the ion focus region and reach the grain. For subsonic regime, $M=0.5$ (see subplot(c) Fig.~\ref{fig:Figure2}), we see the potential is ``bent" towards the grain and is qualitatively similar to the  results observed in LR calculations~\cite{Joost:PPCF2015}. However, for higher ion streaming speeds, i.e. $M=1.0$, we observe that the potential is somehow compressed onto the streaming axis and the wake potential contour along the direction of flow is bell shaped (see subplot(h)  Fig.~\ref{fig:Figure2}) with very small density of ions in the transverse direction near the grain.
In the supersonic regime i.e. $M=1.5$ increasing the strength of magnetization, cf.  Fig.~\ref{fig:Figure2} subplot (i),  leads to better penetration of ion focus region by electrons eventually leading to stronger streaming of electrons upstream towards grain. The potential contour is no longer  limited to bell shape rather like a flattened rod completely wiping out the ion focus in the vicinity of the grain. 
 
To understand the interplay of streaming speed with magnetization strength on the wakefield profile, let us revisit Fig.~\ref{fig:Figure2} from the perspective of streaming speed. Even a small amount of magnetization initiates the motion of electrons upstream towards the grain and ions are pushed in the transverse direction, at higher Mach numbers. In the unmagnetized case, upstream propagation of any disturbance is suppressed for the supersonic ion speed. However, for the strongly magnetized case with ions flowing at supersonic speed, we see the strong upstream propagation of electrons.
 %
\begin{figure}
\includegraphics[scale=1,trim = 0cm 0cm 0cm 0cm, clip =true]{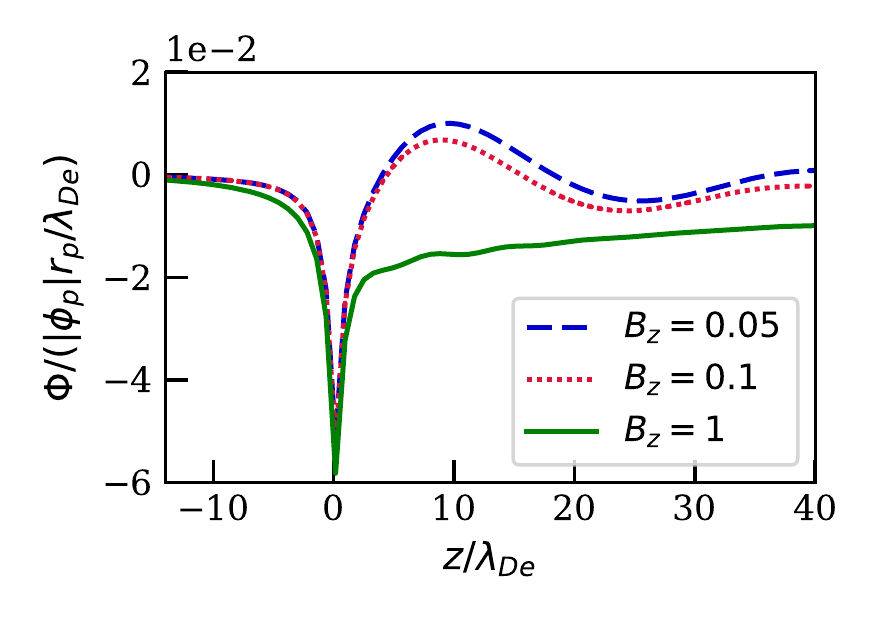}
\caption{Wake potential along the streaming axis  for $M=1$ ($M_{th}=10$) and $B_z=0.05$ (blue dashed line), $B_z=0.1$ (red dotted line), and $B_z=1.0$ (green solid line). 
}
\label{fig:Figure3}
\end{figure}

In Fig.~\ref{fig:Figure3}, the wake potential variation along magnetization (and streaming direction) is shown for three different values of magnetization strength, $\beta=0.1, 0.5$, and $1.0$.  The  grain is at the origin and the normalized grain charge is $\bar{Q}_d= 0.01$. Here, one can observe clearly that the increasing strength of magnetic field decreases the amplitude of wake oscillations downstream grain. At higher streaming speed, when magnetization strength is such that the ion gyro-frequency is equal to or greater than the ion plasma frequency, the wake oscillations positive peak is completely subdued. Note that, it has been envisaged in LR simulations~\cite{Ludwig:NJP2012}, and is not a numerical artifact. The new features that we see now due to higher resolution and contribution of nonlinear features, which has not been observed in LR simulations, is that the electrons are more streamlined and move upstream at higher magnetic field strengths which was forbidden in the case without magnetic field.

\begin{figure}
\includegraphics[scale=1,trim = 0cm 0cm 0cm 0cm, clip =true]{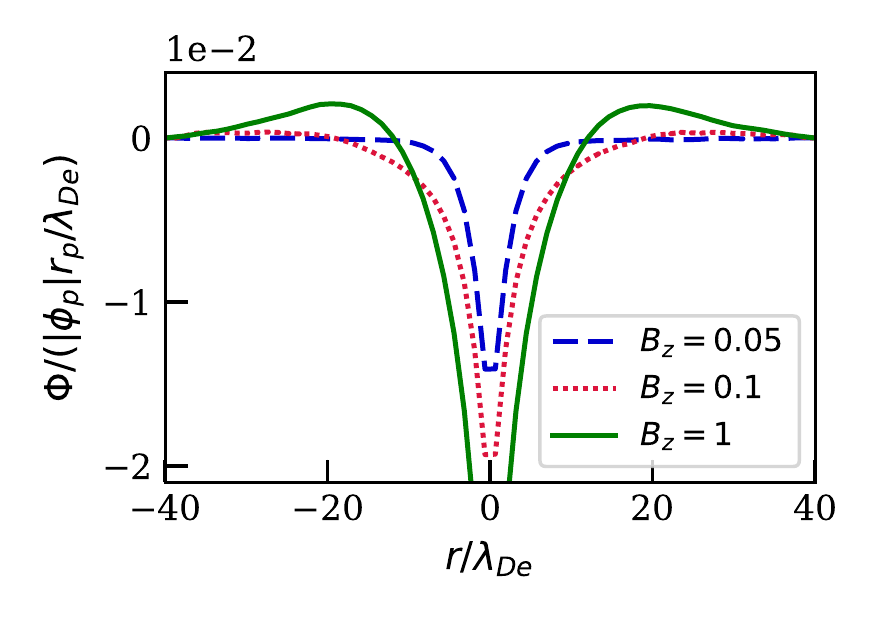}
\caption{Wake potential transverse to the streaming axis  for $M=0.5$ ($M_{th}=5$) and $B_z=0.05$ (blue dashed line), $B_z=0.1$ (red dotted line), and $B_z=1.0$ (green solid line). }
\label{fig:Figure4}
\end{figure}
The wake potential variation in the direction transverse to the magnetization (and ion flow) is shown in Fig.~\ref{fig:Figure4}  for three different values of magnetization strength, $\beta=0.1$, $0.5$, and $1.0$ for ions streaming in subsonic regime, $M=0.5$. Transverse to the streaming direction, we see that with increasing magnetization, oscillation amplitude increases in the subsonic regime. This is due to the fact that the ions streaming in closer proximity of the grain are scattered and an ion depletion is created behind the grain.  Some of these scattered ions assemble in the vicinity of the grain in the transverse direction manifesting transverse oscillations.

\begin{figure}
\includegraphics[scale=1,trim = 0cm 0cm 0cm 0cm, clip =true]{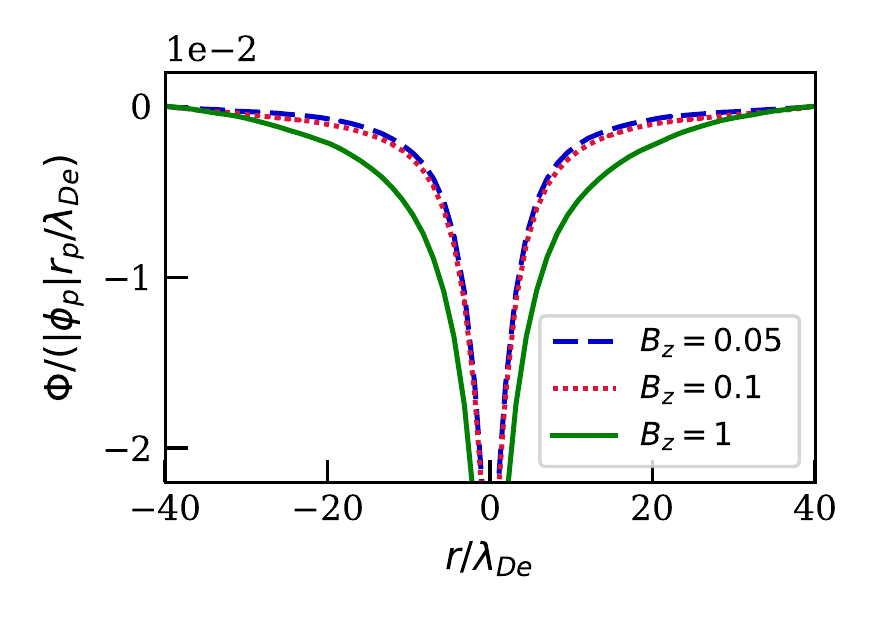}
\caption{Wake potential  transverse to the streaming axis  for $M=1$ ($M_{th}=10$) and $B_z=0.05$ (blue dashed line), $B_z=0.1$ (red dotted line), and $B_z=1.0$ (green solid line).  }
\label{fig:Figure5}
\end{figure}
In Fig.~\ref{fig:Figure5}, the wake potential variation in the direction transverse to the magnetization (and streaming direction) is shown for magnetization strengths, $\beta=0.1, 0.5$, and $1.0$ with ions in the sonic regime, $M=1.0$.
At higher streaming speeds, there is no more accumulation of ions in the transverse direction. Streaming ion flow is strong enough to make the ions overcome the electrostatic potential barrier due to grain and wake potential, and hence is able to sweep them completely from the grain proximity.

 \begin{figure}
 \includegraphics[scale=0.85, trim = 0cm 0.0cm 0cm 0.0cm, clip =true, angle=0]{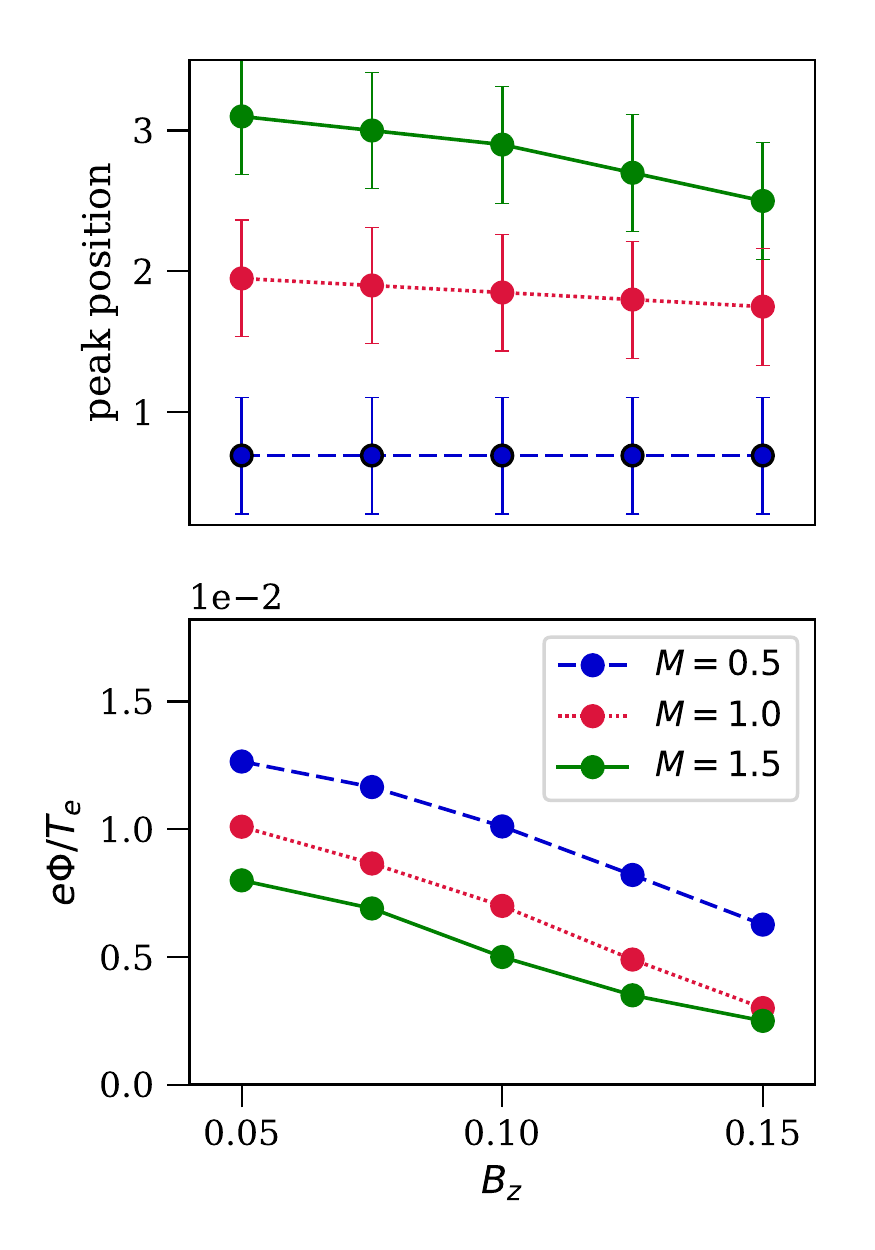} 
\caption{Variation of the maximum of the  peak position (left subplot) and peak amplitudes (right subplot) of the wake potential as a function
of magnetic field strength in normalized units for the shifted Maxwellian distribution at $\nu/\omega_{pi}=0.002$.}
\label{fig:Figure6}
 \end{figure}
 
We present the variations of the maximum peak position and the peak height with magnetization, for various streaming speeds, $M=0.5, 1, 1.5$ in Fig.~\ref{fig:Figure6}.
In accordance with previous reported results by Joost et. al.~\cite{Joost:PPCF2015}, here also, we see a decline in peak position and amplitude of peak height with increasing strength of magnetization. For subsonic regime, ion focusing occurs very near to the grain due to its longer interaction time with grain and smaller kinetic energy. As the streaming speed increases, the ions have higher kinetic energy to move past the grain to a farther location and hence the peak position for higher streaming speeds is farther than the subsonic ion flow. The trend of the peak position with streaming speeds is such that it increases with increasing Mach number.
We observe that the wake potential amplitude for subsonic regime is higher than that for supersonic regime for the case of magnetized ion flow. 
Increasing magnetization strength doesn't change the wake peak position significantly for ions streaming at subsonic speeds. However, we notice that the wake peak position shifts closer to grain with increasing magnetization strength for supersonic ion flows.

\subsection{Impact of magnetization strength on the ion density distribution}
\begin{figure*}
 \includegraphics[scale=0.68, trim = 0cm 0cm 0cm 0cm, clip =true, angle=0]{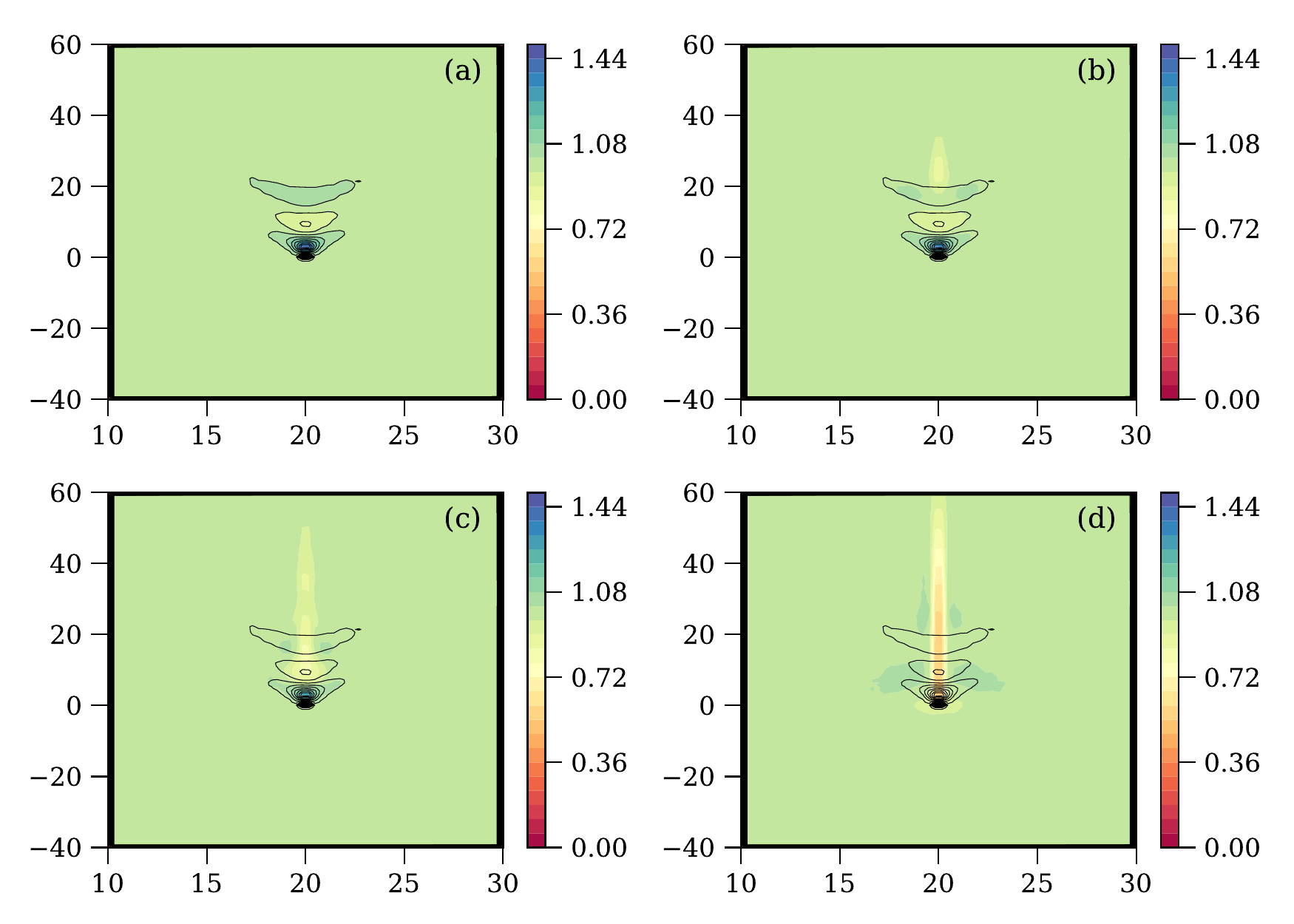} 
\caption{Spatial profiles of the ion density (normalized to the distant unperturbed ion density), averaged over the azimuthal angle, for various strengths of magnetic field : (a) $\beta=0.05$, (b) $\beta=0.1$,  (c) $\beta=0.15$,  and (d) $\beta=1.0$ with streaming velocity $M=0.5$.
}
\label{fig:Figure7}
 \end{figure*} 
In Fig.~\ref{fig:Figure7}, we present the density contour for various magnetizing strengths with streaming speed $M=0.5$. For very small magnetic field strength, it exhibits the pattern as one observed in the unmagnetized ion case~\cite{Sundar:POP2017, Ludwig:NJP2012}. As   the strength of the magnetic field increases, the density downstream grain develops as candle flame structure similar to the induced density density distribution reported by Zhandos et.al.~\cite{Zhandos:arxiv2017}. Further increase in the strength of the field leads to elongation and broadening  of the candle flame shaped ion density distribution downstream grain. Basically, what we see as  elongated structure is the ion depletion created behind grain due to the strong magnetic field applied along flow.

 \begin{figure*}
 \includegraphics[scale=0.68, trim = 0cm 0cm 0cm 0cm, clip =true, angle=0]{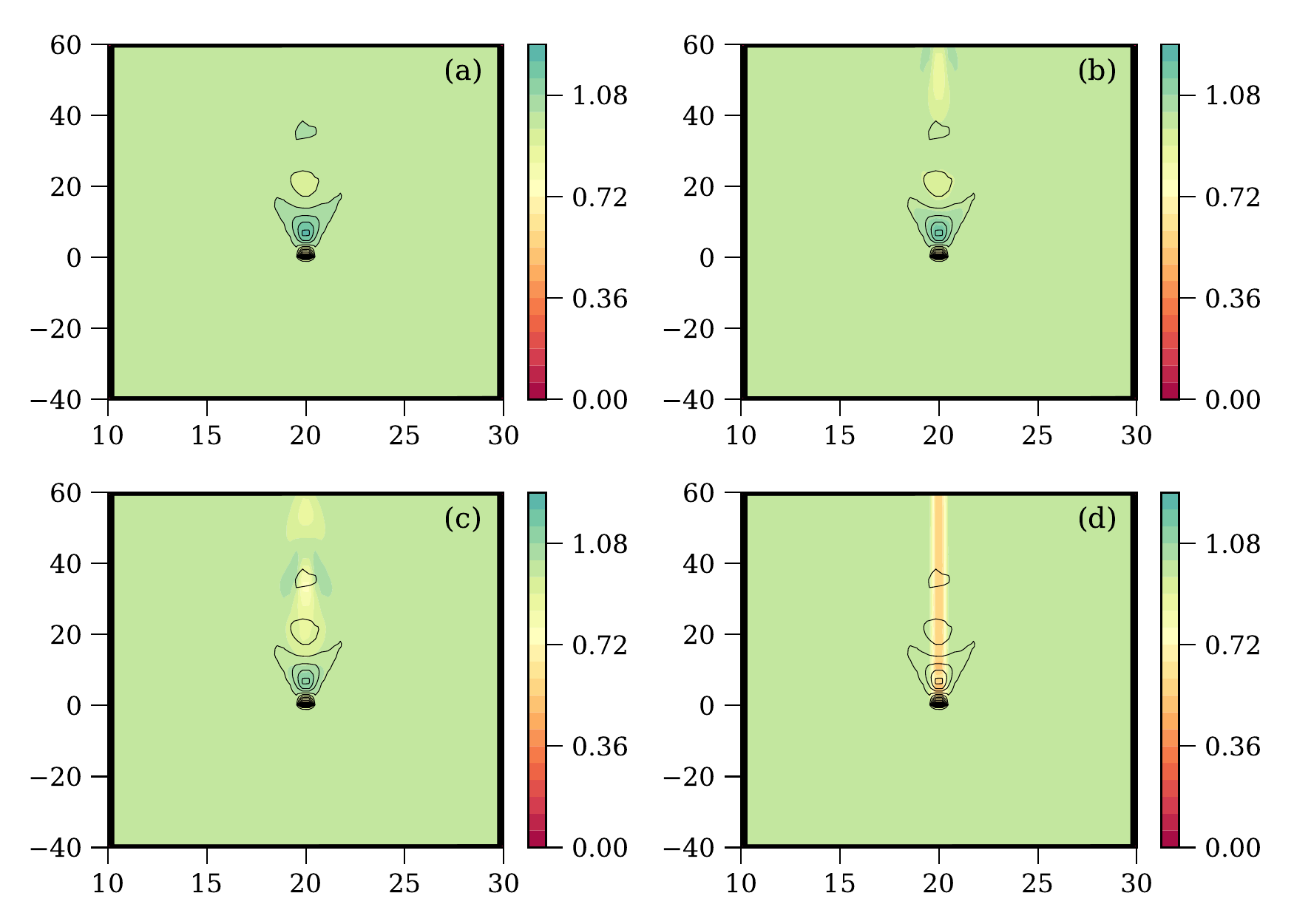} 
\caption{
Spatial profiles of the ion density (normalized to the distant unperturbed ion density), averaged over the azimuthal angle, for various strengths of magnetic field : (a) $\beta=0.05$,  (b) $\beta=0.1$,  (c) $\beta=0.15$,  and (d) $\beta=1.0$ with streaming velocity $M=1.0$.
}
\label{fig:Figure8}
 \end{figure*}
We present the density contour for various magnetizing strengths in the sonic regime $M=1.0$ in Fig.~\ref{fig:Figure8}. For very small magnetic field strength, it exhibits the pattern as one observed in the unmagnetized ion case as usual. However, as one increases the strength of the magnetic field, the density contour exhibits an altogether different pattern. It evolves as low density candle flame structure moving from far end of the wake towards the grain with increasing magnetization strength. Elongation of ion depletion region is stronger in this case. This can be understood with our wake potential  profile explanation wherein we see the streaming of electrons from the far end propagating upstream towards the grain.


\section{Conclusion}

In the present work, we have investigated the electrostatic potential distribution around a point-like charged grain in a streaming plasma in the presence of magnetic field applied along the ion streaming direction for a shifted-Maxwellian ion distribution function.
 We provide here an advanced numerical work by including the effect of an external electric and magnetic field with ion-neutral charge-exchange collisions on the wake potential and ion density distribution. The electric field imitates  the sheath region of electric discharges where the electric field makes the the ions accelerate towards the electrodes. 
 The presence of magnetic field suppresses the wake amplitude formed downstream grain. Impact of streaming ion speed on these magnetized ion wake manifests in the wake flown away farther from the grain corresponding to the applied streaming speed strength. In the presence of magnetic field, even the  low-velocity ions alter  ion focusing behind the dust grain eventually modifying the the wake potential substantially. 
       
Using LR formalism, the study of the impact of magnetization on wake was attempted by Joost et. al.~\cite{Ludwig:NJP2012}. They discussed the limitations of the LR formalism regarding non-Maxwellian distribution and apparent symmetry breaking in the presence of neutrals. For small streaming speeds i.e. upto $M\sim 0.5$ our result is qualitatively similar to the results presented in the paper by Joost et. al.~\cite{Ludwig:NJP2012}. Our results also indicate that the potential is somewhat ``bent" towards the grain at high beta (Fig. 3). For $M>1$ the potential is somehow compressed onto the streaming axis. This resembles closely to the observations made by Joost et. al.~\cite{Ludwig:NJP2012}. 
However, there are few striking differences in the characteristic profile of wake potential which has its origin in the difference in the basic system configuration adopted in the two cases. As mentioned in the Section I, we have,  a stationary grain in streaming magnetized ions. On the other hand, Joost et. al.~\cite{Joost:PPCF2015} considered the case for grain moving in stationary plasma.

																	An important observation is damping of the wake potential with the magnetic field strength for the entire flow range and many novel features in the wake and density downstream grain.  The location of the wake peak maximum exhibited strong dependence on the Mach number as well as magnetic field strength. 
  For the smaller streaming speeds, as the magnetic field strength increases, we see a depletion in the ion charge density at the far end downstream grain which eventually facilitates the electrons to stream through the depleted positive space charge region towards the grain.  Upon further increase in  the streaming speed,  electron upstream flow is more pronounced and manifest in the form of bell-shape negative potential around grain wiping out the ion focus from the dust vicinity. 
The electrons begin to streamline for comparatively smaller value of magnetization. Nevertheless, the upstream propagation of electrons is observed to be significantly predominant for supersonic ion flows in stronger magnetic field. We have described  in detail a fairly qualitative physics for the grain-wakefield phenomena for streaming magnetized ions which will contribute to the understanding of inter-grain interaction for magnetized dusty plasmas and render assistance in revealing underlying physics in future.


\section{Acknowledgments}
S. Sundar would like to thank I. H. Hutchinson for support in using the COPTIC code and acknowledge support of CAU Kiel. This work was supported by the DFG via SFB-TR24, project A9. Our numerical simulations were performed
at the HPC cluster of Christian-Albrechts-Universit{\"a}t zu Kiel.
S. Sundar would also like to thank M. Bonitz, H. K{\"a}hlert,  P. Ludwig, J.-P. Joost, and  Dr. Zh. Moldabekov for their help in scientific discussion.



%

\end{document}